\documentclass[twocolumn,showpacs,preprintnumbers,amsmath,amssymb,superscriptaddress,nofootinbib,english]{revtex4-1}
\usepackage{times,amsmath,amsfonts,amssymb,epstopdf}
\usepackage{graphicx}
\usepackage{dcolumn}
\usepackage{bm}
\usepackage{epsfig}
\usepackage{graphicx}
\usepackage{hyperref}
\usepackage[usenames]{color}
\usepackage{url}
\usepackage[normalem]{ulem}
\usepackage[T1]{fontenc}
\usepackage[dvipsnames]{xcolor}

\def\be{\begin{equation}}
\def\ee{\end{equation}}
\def\ba{\begin{eqnarray}}
\def\ea{\end{eqnarray}}
\begin{document}

\title{The Layzer-Irvine Equation for Scalar-Tensor Theories: \\A Test of Modified Gravity N-body Simulations}

\author{Hans~A.~Winther}
\email[Email address: ]{h.a.winther@astro.uio.no}
\affiliation{Institute of Theoretical Astrophysics, University of Oslo, 0315 Oslo, Norway}

\date{\today}

\begin{abstract}
The Layzer-Irvine equation describes energy conservation for a pressure less fluid interacting though quasi-Newtonian gravity in an expanding Universe. We here derive a Layzer-Irvine equation for scalar field theories where the scalar field is coupled to the matter fields, and show applications of this equation by applying it to N-body simulations of modified gravity theories. There it can be used as both a dynamical test of the accuracy of the solution and the numerical implementation when solving the equation of motion. We also present an equation that can be used as a new static test for an arbitrary matter distribution. This allows us to test the N-body scalar field solver using a matter distribution which resembles what we actually encounter in numerical simulations.
\end{abstract}


\maketitle

\section{Introduction}

The apparent accelerated expansion of the Universe \cite{au1,au2} is one of the biggest puzzles in modern cosmology. There exist several theoretical explanations for it and these generally goes under the broad term dark energy \cite{dereview}. 

Dark energy in the form of a cosmological constant is currently the best fit to observations, but it has several theoretical problems like the fine-tuning and the coincidence problem. Some of these problems can be alleviated if the energy density of the cosmological constant becomes dynamical. This approach leads to dark energy models where the accelerated expansion is due to some new dynamical field \cite{cosmon}. The dark energy field(s) evolves on cosmological time scales, and therefore if dark energy has interactions with ordinary baryonic matter then a cosmologically long range fifth-force will be the result \cite{carroll}. 

Gravity is very well tested in the solar system and the results agree perfectly with the predictions of General Relativity \cite{will}.  A gravitational interaction that differ from General Relativity are at odds with local gravity experiments and in models where the dark energy is coupled to dark matter (like coupled quintessence \cite{cq}) it is therefore generally assumed that there is no coupling to baryons. If a coupling to baryons do exist (we call this scenario modified gravity) then a screening mechanism \cite{screen} is required to evade local experiments and at the same time give rise to interesting dynamics on cosmological scales.

In the last decade several modified gravity models with a screening mechanism, most based on a single scalar degree of freedom, have been put forward. Models following from works on massive gravity such as DGP \cite{dgpm} and the Galileon \cite{gal1,gal2} are well known examples. Another class of models are the chameleon-like models such as the chameleon/$f(R)$ \cite{chameleon,mota_shaw,chameleon_cosmology,chameleon_winther}, symmetron \cite{symmetron,symmetron_olive} and environmental dependent dilaton \cite{env_dilaton}.

For this last class of models it has been shown that the background cosmology is generally very close to that of $\Lambda$CDM. However, even though the background cosmology is the same, the growth of linear perturbations is modified and alters structure formation. One can also show quite generally that the results of local gravity experiments implies a interaction range in the cosmological background today in the sub megaparsec region \cite{unified_winther}. This is in the range where perturbations in the fiducial $\Lambda$CDM model goes from being well described by linear theory to where one needs more elaborated methods like N-body simulations to make accurate predictions of the theory.

N-body simulations for modified gravity theories require one to fully solve for the 3D distribution of the scalar field just as one normally does for the gravitational potential. The highly non-linear form of the field-equation makes this computationally challenging. Recently, several different N-body codes have been created that do this job \cite{ecosmog,isis,mgadget,chicago_code,mlapm,mota_claudio}, and studies of structure formation in the non-linear regime have been performed for many different modified gravity models like for example the chameleon$/f(R)$ gravity \cite{chicago_code2,fofr_baojiu,syst2_winther,baldifofr}, the symmetron \cite{symmetron_winther,syst1_winther}, the environmental dependent dilaton \cite{dilaton_nbody}, the DGP model \cite{dgp_code,baojiu_dgp} and phenomenological fifth-force models \cite{hell}. For a review of N-body simulations for non-standard scenarios see \cite{baldi1}.

One important lesson learned from these studies is that one needs simulations to make accurate predictions: linear perturbation theory gives inaccurate results for almost all scales where the matter power-spectra differs from $\Lambda$CDM \cite{syst1_winther,syst2_winther}.

Before performing such simulations the scalar field solver needs to be properly tested for both static and dynamical cases where analytical or semi-analytical solutions exist. For the static case several tests already exist \cite{ecosmog}, while for the time evolution of the cosmological simulations so far the only real test is to compare the results with that of other codes.

There is however one other test based on energy conservation, that so far has been ignored for modified gravity simulations, which can be used for this purpose. For collisionless N-body simulation (i.e. dark matter only simulations) a Newtonian energy conservation equation, taking into account the expanding background, exist and is known as the Layzer-Irvine equation \cite{layzer,irvine}. This equation gives a relation between the kinetic energy and the gravitational potential energy of dark matter particles and is valid throughout the process of structure formation. The equation only applies for standard gravity and needs to be generalized if we want to use it for modified gravity theories. 

The idea to look at extensions and generalizations of the Layzer-Irvine equation for models beyond $\Lambda$CDM is not new. In \cite{lieq_1}, the equation was extended to a dark energy component with an arbitrary equation of state and then generalized to account for a non-minimal interaction between dark matter and dark energy. The spherical collapse model was applied in \cite{lieq_2} to derive a generalized Layzer-Irvine equation for the case where the dark energy can cluster and used to estimate the maximum impact that dark energy perturbations can have on the dynamics of clusters of galaxies. A Layzer-Irvine equation for interacting dark energy models was derived in \cite{lieq_new,lieq_wang}, using perturbation theory, and then applied to study how dark matter and dark energy virializes. In \cite{lieq_4} the equation was derived for several phenomenological gravitational force-laws. The equation have also been applied to observations to put constraints on the coupling between dark matter and dark energy \cite{lieq_new}.

In this paper, we derive the Layzer-Irvine equation for a quite general class of modified gravity models and the methods we use can easily be extended to any scalar field model of interest. We implement the resulting equation in an N-body code and show that it can be used as a new dynamical test for N-body codes of modified gravity.

The setup of this paper is as follows. We begin by briefly reviewing scalar-tensor theories of modified gravity in Sec.~(\ref{sec1}) and the Layzer-Irvine equation for standard gravity in Sec.~(\ref{sec2}).  The modified Layzer-Irvine equation is derived in Sec.~(\ref{sec3}) and we discuss how to implement this equation in an N-body code in Sec.~(\ref{sec4}). In Sec.~(\ref{sec5}) we present the results from tests on N-body simulations of modified gravity before we summarize and conclude in Sec.~(\ref{sec6}).

Throughout this paper we use units of $c=\hbar=1$ and the metric signature $(-,+,+,+)$.

\section{Scalar-tensor theories of Modified Gravity}\label{sec1}
In this section be briefly review scalar-tensor modified gravity theories. We are in this paper mainly interested in scalar-tensor theories defined by the action
\begin{align}\label{action}
S = &\int d^4x\sqrt{-g}\left[\frac{R}{16\pi G} + f(X,\phi)\right] \nonumber\\
&+ S_m(A^2(\phi)g_{\mu\nu};\psi_m)
\end{align}
where $R$ is the Ricci scalar, $G$ is the bare gravitational constant, $g$ is the determinant of the metric $g_{\mu\nu}$, $\phi$ the scalar field, $X = -\frac{1}{2}g^{\mu\nu}\phi_{,\mu}\phi_{,\nu}$ and $\psi_m$ represents the different matter-fields which are coupled to the scalar field $\phi$ via the conformal rescaled metric $\tilde{g}_{\mu\nu}=A^2(\phi)g_{\mu\nu}$. 

The Einstein equations follows from a variation of the action with respect to $g_{\mu\nu}$ and reads
\begin{align}\label{eeq}
R_{\mu\nu} - \frac{1}{2}Rg_{\mu\nu} = 8\pi G\left[A(\phi)T^m_{\mu\nu} + T^\phi_{\mu\nu}\right]
\end{align}
where $T^m_{\mu\nu}$ is the energy-momentum tensor for the matter fields and 
\begin{align}\label{emphi}
T^{\phi}_{\mu\nu} = f_X\phi_{,\mu}\phi_{,\nu} + g_{\mu\nu}f,~~~~~~~~f_X \equiv \frac{\partial f}{\partial X}
\end{align}
is the energy-momentum tensor for the scalar field.

The Klein-Gordon equation for $\phi$ follows from a variation of the action with respect to $\phi$ and reads
\begin{align}\label{eomphi}
\nabla_{\mu}(f_X\nabla^{\mu}\phi) = - f_{,\phi} - A_{,\phi} T_m
\end{align}
where $T_m = g^{\mu\nu}T^{m}_{\mu\nu}$ is the trace of the energy-momentum tensor of the matter field(s). In the rest of this paper we will only consider a single dust like matter component for which $T^m = -\rho_m$. The conformal coupling of $\phi$ to matter gives rise to a fifth-force which in the non relativistic limit and per unit mass is given by
\begin{align}
\vec{F}_\phi = -\vec{\nabla}\log A = -\frac{\beta(\phi)}{M_{\rm Pl}}\vec{\nabla}\phi,~~~~\beta(\phi) \equiv M_{\rm Pl}\frac{d\log A(\phi)}{d\phi}
\end{align}
The Bianchi identity and the field equations implies the following conservation equations
\begin{align}\label{conseq}
\nabla_{\mu}T_\phi^{\mu\nu} &= +\frac{\partial\log A}{\partial \phi}A(\phi) T_m^{\mu\nu}\nabla_{\mu}\phi\\
\nabla_{\mu}(A(\phi)T_m^{\mu\nu}) &= -\frac{\partial\log A}{\partial \phi}A(\phi) T_m^{\mu\nu}\nabla_{\mu}\phi\\
\nabla_{\mu}T_m^{\mu\nu} &= 0
\end{align}
The equations presented above are the only ones needed to derive the modified Layzer-Irvine equation. For a more thorough review of scalar tensor modified gravity theories see \cite{mgreview}.

\section{The Layzer-Irvine equation for General Relativity}\label{sec2}
In this section we re-derive the Layzer-Irvine equation for the case of a collisionless fluid interacting with gravity in an expanding background. This equation was first derived by Layzer \cite{layzer} and Irvine \cite{irvine} in the early 1960s and our derivation below will be close up to that of \cite{layzer}.
\\\\
We will here only consider a flat spacetime. However, the results we derive below also applies for curved spacetimes as long as we only apply them to regions smaller than the radius of curvature \cite{layzer}. The background metric of a flat homogenous and isotropic Universe is the Friedmann-Lema$\hat{\text{i}}$tre-Robertson-Walker metric
\begin{align}
ds^2 = -dt^2 + dr^2 = -dt^2 + a^2(t)(dx^2+dy^2+dz^2)
\end{align}
In the following $\vec{x}$ will denote the comoving coordinate and $\vec{r}=a\vec{x}$ the physical coordinate. For a collection of collisionless particles the energy momentum tensor is given by
\begin{align}
T_m^{\mu\nu}(\vec{x'}) = \sum_i \frac{m_i\delta(\vec{x'}-\vec{x_i})u_i^{\mu}u_i^{\nu}}{\sqrt{-g}}
\end{align}
where $u_i^{\mu}$ is the four velocity of particle $i$. If we treat the collection of particles as a fluid then we can define 
\begin{align}
T_m^{\mu\nu} = \rho_m u^{\mu}u^{\nu}
\end{align}
where $u^{\mu}$ is the four-velocity of the fluid. We let $\rho_m(r,t) = \overline{\rho}_m(t) + \delta\rho_m(r,t)$ denote the matter density field and $\vec{v} = a\dot{\vec{x}}$ the peculiar velocity field.  An overbar will always denote a quantity defined in the background cosmology, e.g. $\overline{\rho}_m(t)$ is the homogenous and isotropic component of the matter field.

The continuity equation for the energy-momentum tensor reads
\begin{align}
\nabla_{\mu} T_m^{0\mu} = 0~~~~\to~~~~~\nabla_{\mu}(\rho_m u^{\mu}) = 0
\end{align}
By writing out the components and subtracting off the background equation, $\dot{\overline{\rho}}_m+3H\overline{\rho}_m=0$, we get it on a convenient form
\begin{align}\label{conteq}
\dot{(a^3 \delta\rho_m)} + a^3\vec{\nabla}_r(\rho_m \vec{v})=0
\end{align}
In the real Universe the metric is perturbed due to the presence of matter perturbations and this equation will have additional contributions like terms containing the time derivative of the Newtonian potential $\Phi_N$. These terms can generally be neglected as long as the weak-field approximation $\Phi_N\ll 1$ holds (which is the case for most cosmological and astrophysical applications).

The equation describing the motion of the particles (fluid) is the geodesic (Euler) equation, 
\begin{align}
\frac{du^{i}}{d\tau} + \Gamma^{i}_{\mu\nu}u^{\mu}u^{\nu}=0,~~~~~~~~~~u^i = \frac{dx^i}{d\tau}
\end{align}
If we take the energy-momentum tensor of matter to be that of particles then this equation follows directly from the Bianchi identity. Writing out the geodesic equation and neglecting small terms, we get an equation of motion very similar to the Newtonian result generalized to an expanding background
\begin{align}
\ddot{\vec{x}}+2H\dot{\vec{x}} = -\frac{1}{a}\vec{\nabla}_r \Phi_N
\end{align}
or equivalently
\begin{align}\label{eq1}
\frac{\partial(a\vec{v})}{\partial t} = -\vec{\nabla}_r (a\Phi_N)~~~~~~~~v^i = a \dot{x}^i
\end{align}
The Newtonian gravitational potential is determined by the Poisson equation
\begin{align}\label{eq2}
\nabla_r^2\Phi_N = 4\pi G\delta\rho_m
\end{align}
and the solution can also be written explicitly as
\begin{align}
\Phi_N(r,t) = -G\int\frac{\delta\rho_m(r',t) d^3r'}{|r-r'|}
\end{align}
where the integration is over the whole space. The system of equations
\begin{align}
\ddot{\vec{x}}+2H\dot{\vec{x}} &= -\frac{1}{a}\vec{\nabla}_r \Phi_N\\
\nabla_r^2\Phi_N &= 4\pi G\delta\rho_m
\end{align}
forms the basis of N-body simulations for collisionless matter. 

To form the Layzer-Irvine equation we need to integrate the equation of motion Eq.~(\ref{eq1}) over space. In the following we will consider a very large, but finite, volume to be able to neglect surface terms arising from integration by parts and to avoid convergence problems. It is also possible to consider, as is the case for N-body simulations, a finite volume with periodic boundary conditions. We will in the next section discuss how to handle the case of going to an infinite volume, which turns out to be pretty straightforward and does not change the form of the final equation.

To form the Layzer-Irvine equation we contract Eq.~(\ref{eq1}) with $\vec{v}a\rho_m d^3r = \vec{v}a^4\rho_m d^3x$ and integrate over the distribution of particles with the result
\begin{align}\label{eq3}
\frac{\partial T}{\partial t} + 2HT = -\int d^3r (\rho_m\vec{v})\cdot(\vec{\nabla}_r\Phi_N) 
\end{align}
where 
\begin{align}
T = \int \frac{1}{2}v^2\rho_m d^3r = \sum_{i=1}^{N_{\rm particles}}\frac{1}{2}m_i v_i^2
\end{align}
denotes the total kinetic energy associated with the peculiar motion. Using integration by parts and applying the continuity equation Eq.~(\ref{conteq}) we can rewrite the right hand side of Eq.~(\ref{eq3}) as
\begin{align}
-\int (\vec{\nabla}_r\Phi_N)\cdot \vec{v} \rho_m d^3r &= \int \Phi_N \vec{\nabla}_r(\vec{v} \rho_m) d^3r\nonumber\\
&= -\int \Phi_N \frac{\partial}{\partial t}(\delta\rho_m d^3r)
\end{align}
which can be rewritten once again using the Poisson equation as
\begin{align}
-\int \Phi_N \frac{\partial}{\partial t}(\delta\rho_m d^3r) &= -\left(\frac{\partial U_N}{\partial t} + HU_N\right)
\end{align}
where 
\begin{align}
U_N &= \int \frac{1}{2}\Phi_N \delta\rho_m d^3r \nonumber\\
&= -\frac{G}{2}\int\int \frac{\delta\rho_m(r,t)\delta\rho_m(r',t)d^3rd^3r'}{|r-r'|}
\end{align}
is the gravitational potential energy. Collecting results we are left with
\begin{align}\label{lieq}
\frac{\partial}{\partial t}\left(T + U_N\right) + H(2T+ U_N) = 0
\end{align}
which is the Layzer-Irvine equation.

If the total energy $E=T+U_N$ is conserved we recover the well known virial relation $2T+U_N = 0$.

By making the definitions (the justifications for these definitions in terms of statistical physics of fluids have been given by Irvine \cite{irvine})
\begin{align}
 \epsilon_m &= \frac{T+U_N}{\mathcal{V}}\\
3p_m &= \frac{2T+U_N}{\mathcal{V}}
\end{align}
where\footnote{For an infinite volume this is to be understood as a limiting procedure.} $\mathcal{V}=\int d^3r$ we have that Eq.~(\ref{lieq}) can be written on the more familiar form
\begin{align}
\frac{\partial}{\partial t}\epsilon_m + 3H(\epsilon_m+ p_m) = 0
\end{align}
which is a cosmological continuity equation.

\section{Layzer-Irvine equation for Scalar-Tensor theories}\label{sec3}
In this section we derive the Layzer-Irvine equation for the class of scalar-tensor (modified gravity) theories given by the action Eq.~(\ref{action}). We will just state the equations describing our system without derivation, as a complete derivation of the equations below can be found in e.g. \cite{dilaton_nbody}.

As we did in the previous section we take the energy-momentum tensor of the matter to be that of particles. Note that we use the definition of $T_m^{\mu\nu}$ depicted in Eq.~(\ref{eeq}) so that  the density $\rho_m$ satisfies the usually continuity equation Eq.~(\ref{conteq}), but as we will see below the Newtonian potential is sourced by the density $\rho_J \equiv A(\phi)\rho_m$. The continuity equation in terms of this density reads
\begin{align}\label{contmod}
\frac{\dot{(a^3 \delta\rho_J)}}{a^3} &+ \vec{\nabla}_r(\rho_J \vec{v}) - \rho_J \vec{v}\vec{\nabla}_r \log A\nonumber\\
&-\dot{\log A}\delta\rho_J -  \overline{\rho}_J \frac{\partial}{\partial t}\log \frac{A}{\overline{A}}=0
\end{align}
where $\delta\rho_J = A(\phi)\rho_m - A(\overline{\phi})\overline{\rho}_m$.

The geodesic equation describing the motion of the fluid is modified due to the presence of the coupling of $\phi$ to matter
\begin{align}
\frac{du^{i}}{d\tau} + \Gamma^{i}_{\mu\nu}u^{\mu}u^{\nu} = -\frac{d\log A}{d\phi}\left(\phi^{,i} + u^{\mu}\phi_{,\mu}u^i\right)
\end{align}
which in the non-relativistic limit becomes
\begin{align}\label{geomod}
\frac{\partial}{\partial t}(a\vec{v}) + (a\vec{v})\frac{\partial \log A}{\partial t} = -a\vec{\nabla}_r(\Phi_N+\log A)
\end{align}
The Poisson equation is also modified due to the presence of the scalar field and reads 
\begin{align}\label{poimod}
\nabla_r^2\Phi_N = 4\pi G\delta\rho_J + 4\pi G\delta S_\phi \equiv 4\pi G\delta S_{\rm tot}
\end{align}
where the source coming from the scalar field is
\begin{align}
\delta S_\phi = \delta \rho_\phi + 3\delta p_\phi
\end{align}
with $\delta \rho_\phi = \rho_\phi - \overline{\rho}_\phi$ and likewise for $\delta p_\phi$. The energy density and pressure of the scalar field is defined as $\rho_\phi = T^0_{\phi~0}$ and $p_\phi = \frac{1}{3}T^i_{\phi~i}$ respectively.
\\\\
Contracting Eq.~(\ref{geomod}) with $a\vec{v}\rho_Jd^3r=a^4\vec{v}\rho_Jd^3x$ and integrating up we find
\begin{align}\label{eq5}
\dot{T} + H(2T+\delta T) = -\int \vec{\nabla}_r(\Phi_N + \log A)\rho_J\vec{v}d^3r
\end{align}
where
\begin{align}
T &= \int d^3r \frac{1}{2}v^2\rho_J\\
\delta T &= \int d^3r \frac{1}{2}v^2\rho_J \left(\frac{\partial \log A}{\partial \log a}\right)
\end{align}
Using the continuity equation Eq.~(\ref{contmod}) we can remove the velocity term in Eq.~(\ref{eq5}) by integration by parts to find
\begin{align}\label{me1}
&\int \vec{\nabla}_r(\Phi_N + \log A) \vec{v} \rho_J d^3r = \\
\label{A1}&+\int \Phi_N\left(\frac{\partial}{\partial t}(\delta S_{\rm tot} d^3r)\right)\\
\label{A2}&- \int \Phi_N\left(\frac{\partial}{\partial t}(\delta S_\phi d^3r)\right)\\
\label{A3}&+\int \log A\left(\frac{\partial}{\partial t}(\delta\rho_J d^3r)\right)\\
\label{A4}&- \int d^3r (\Phi_N + \log A)\delta S_{\rm tot}\frac{\partial\log A}{\partial t}\\
\label{A5}&+ \int d^3r (\Phi_N + \log A)\delta S_\phi\frac{\partial\log A}{\partial t}\\
\label{A6}&- \int d^3r (\Phi_N + \log A)\overline{\rho}_J \frac{\partial}{\partial t}\log \frac{A}{\overline{A}}\\
\label{A7}&- \int d^3r (\Phi_N + \log A) (\vec{\nabla}_r\log A)\rho_J\vec{v}
\end{align}
We will now go through the different terms one by one.

The first term Eq.~(\ref{A1}) can be integrated by parts with the result
\begin{align}\label{me2}
&\int \Phi_N\left(\frac{\partial}{\partial t}(\delta S_{\rm tot} d^3r)\right) = \dot{U}_N + HU_N\\
&U_N = \int \frac{\Phi_N}{2}\delta S_{\rm tot} d^3r = -\frac{1}{8\pi G}\int d^3r (\vec{\nabla}_r\Phi_N)^2
\end{align}
This last form of $U_N$ follows from the Poisson equation and integration by parts and is identical to that of standard gravity except here the Newtonian potential is also sourced by the scalar field.
\\\\
The term Eq.~(\ref{A2}) is of order $\dot{U}_{S_\phi}$ where
\begin{align}
U_{S_\phi} = \int \frac{\Phi_N}{2}\delta S_\phi d^3r
\end{align}
This term cannot be written on a form that does not include time-derivatives of the Newtonian potential\footnote{This is crucial when we later will implement these equations in an N-body code as time-derivatives of the gravitational potential is in most codes not known.}. We will therefore assume $|U_{S_\phi}| \ll |U_N|$ so that we can neglect this term and the term in Eq.~(\ref{A5}). For known modified gravity theories this assumption is usually satisfied (see e.g. \cite{symmetron_winther}).
\\\\
The term Eq.~(\ref{A4}) becomes $-H(2\delta U_N + \delta U_{\log A})$ where
\begin{align}
\delta U_N &= \int d^3r \frac{\Phi_N}{2} \delta S_{\rm tot}\frac{\partial\log A}{\partial \log a}\\
\delta U_{\log A} &= \int d^3r\log A \delta S_{\rm tot}\frac{\partial\log A}{\partial \log a}
\end{align}
In the following all terms $\delta U_x$ will mean $U_x$ with the inclusion of a factor $\frac{\partial\log A}{\partial \log a}$ in the integrand. We have, for example,
\begin{align}
U_N + \delta U_N &= \int d^3r \frac{\Phi_N}{2} \delta S_{\rm tot}\left(1 + \frac{\partial\log A}{\partial \log a}\right)
\end{align}
and similar for all other terms $U_x$ so that all the terms $\delta U_x$ can be neglected when $\left|\frac{\partial\log A}{\partial \log a}\right| \ll 1$.
\\\\
The term Eq.~(\ref{A6}) can be neglected as its a factor $|\Phi_N + \log A| \ll 1$ smaller than a term coming from Eq.~(\ref{A4}) as we will show below.
\\\\
The term Eq.~(\ref{A7}) can also be neglected for most models of interest. To see this, take the "worst-case" scenario of a scalar fifth-force which is proportional to gravity everywhere with some constant strength $\beta$. For this case this term is of order
\begin{align}
2\beta^2(1+2\beta^2)\frac{\partial}{\partial t}\int d^3r \frac{\Phi_N^2}{2}\delta S_{\rm tot}
\end{align}
and the integrand is a factor $2\beta^2(1+2\beta^2)\Phi_N \ll 1$ smaller than the integrand of $U_N$ for the interesting case $\beta \lesssim \mathcal{O}(1)$.
\\\\
The only term left to evaluate is Eq.~(\ref{A3}). The equation needed to rewrite this term can be found by either using the field equation or more directly by using the conservation equation for the energy-momentum tensor of the scalar field Eq.~(\ref{emphi}). For the first approach we start with the field equation
\begin{align}
\mathcal{L}_\phi \equiv &\frac{1}{a^3}\frac{\partial}{\partial t}\left(a^3 f_X\dot{\phi}\right) - \vec{\nabla}_r\cdot\left(f_X\vec{\nabla}_r\phi\right)\nonumber\\
& - f_{,\phi} + \log A_{,\phi}\rho_J = 0
\end{align}
At the background level this equation simplifies to
\begin{align}
\mathcal{L}_{\overline{\phi}} \equiv &\frac{1}{a^3}\frac{\partial}{\partial t}\left(a^3 f_{\overline{X}}\dot{\overline{\phi}}\right) - f_{,\overline{\phi}} + \log A_{,\overline{\phi}}\overline{\rho}_J = 0
\end{align}
The two equations above (trivially) implies
\begin{align}\label{leq}
\int d^3r (\mathcal{L}_\phi\dot{\phi}- \mathcal{L}_{\overline{\phi}}\dot{\overline{\phi}}) = 0
\end{align}
which can be written out and integrated by parts to get it on a convenient form. This procedure applies for any scalar field theory.

The second approach is to start directly from the conservation equation for the scalar field Eq.~(\ref{conseq}) and integrate it over space to get
\begin{align}\label{ueq}
&\frac{\partial}{\partial t}\int d^3r \left(T^{~~0}_{\phi~0} -\overline{T}^{~~0}_{\phi~0}\right) + H\int d^3r (T^{~~i}_{\phi~i}-\overline{T}^{~~i}_{\phi~i})\nonumber\\
&= \int d^3r \left(A(\phi)T_m \frac{\partial \log A}{\partial t}-A(\overline{\phi})\overline{T}_m \frac{\partial \log \overline{A}}{\partial t}\right)
\end{align}
where an overbar as usual denotes a background quantity. This expression is valid for any scalar-field theory in which $f = f(\phi,\partial\phi,\partial\partial\phi,...)$ and not just for our particular $f = f(X,\phi)$. However, if we have a theory where the coupling to the matter sector is not conformal, then the right hand side of this equation needs to be modified.
\\\\
When we specialize to theories given by the action Eq.~(\ref{action}) we find
\begin{align}
&\left(\dot{U}_{\nabla\phi}-HU_{\nabla\phi}\right) + \left(\dot{U}_{\dot{\phi}}+3HU_{\dot{\phi}}\right)+ \left(U_{f} - 3HU_f\right)\nonumber\\
&+ (\dot{U}_A-H\delta U_A)+ \dot{U}_{\log A} = \int \log A\frac{\partial}{\partial t}\left(\delta\rho_Jd^3r\right)
\end{align}
where 
\begin{align}
U_{\nabla\phi} &= \int d^3r f_X\frac{1}{2}(\nabla_r\phi)^2\\
U_{\dot{\phi}} &= \int d^3r f_X\frac{1}{2}\left(\dot{\phi}^2-\dot{\overline{\phi}}^2\right)\\
U_{f} &= \int d^3r \left(g(X,\phi)-g(\overline{X},\overline{\phi})\right)\\
U_A &= \int d^3r \left(\log A(\phi)-\log A(\overline{\phi})\right)\overline{\rho}_J\\
\delta U_A &= \int d^3r \left(\log A(\phi)-\log A(\overline{\phi})\right)\overline{\rho}_J\left(\frac{\partial \log \overline{A}}{\partial \log a}\right)
\end{align}
The $g$ function is defined as $g(X,\phi) \equiv  f_X(X,\phi)X -f(X,\phi)$ and
\begin{align}
U_{\log A} &= \int d^3r \log A\delta S_{\rm tot}\nonumber\\
&= -\frac{1}{4\pi G}\int d^3r \left(\vec{\nabla}_r\Phi_N\right)\cdot(\vec{\nabla}_r\log A)
\end{align}
We can now combine all the results above to get the modified Layzer-Irvine equation
\begin{align}\label{mlieq}
&\frac{\partial}{\partial t}\left(T + U_N + U_{\log A} + U_A + U_{\nabla\phi} + U_f + U_{\dot{\phi}}\right) \nonumber\\
&+ H\left(2T + U_N - U_{\nabla\phi} - 3U_f + 3U_{\dot{\phi}}\right) \nonumber\\
&+ H\left(\delta T - 2\delta U_N - \delta U_{\log A} - \delta U_A\right) = 0
\end{align}
The derivation above assumed a finite volume or a box with periodic boundary conditions. If the volume is infinite we reformulate the equation in terms of
\begin{align}
W_i  = \frac{U_i}{\mathcal{V}}
\end{align}
where $\mathcal{V} = \int d^3r = a^3\int d^3x$. The final equation are then to be read as first integrating over a finite volume $\mathcal{V}$ and then taking the limit $\lim_{\mathcal{V}\to \infty} W_i$. This procedure leaves the equation invariant.

To understand the final equation better we can rewrite it slightly. We start with the space averaged energy density and pressure of the scalar field (the space integral of the $T^{0}_0$ and $T^i_i$ components)
\begin{align}
\epsilon_{\phi} &= \frac{U_{\dot{\phi}}  + U_{\nabla\phi} + U_f}{\mathcal{V}}\\
3p_{\phi} &= \frac{3U_{\dot{\phi}}  - U_{\nabla\phi} - 3U_f}{\mathcal{V}}
\end{align}
We now associate, as we did for standard gravity, 
\begin{align}
\epsilon_m &= \frac{T + U_N}{\mathcal{V}}\\
3p_m &= \frac{2T+ U_N}{\mathcal{V}}
\end{align}
with the internal energy and the cosmic pressure for the matter (due to gravity) and $\epsilon_{\phi m} = \frac{U_{\log A}}{\mathcal{V}}$ with the potential energy associated with the matter-scalar interaction.

Inserting all this in the modified Layzer-Irvine equation, neglecting the (typically) small terms $\delta U_x$, we can write it on the form 
\begin{align}
\frac{\partial}{\partial t}(\epsilon_{\phi} +\epsilon_m + \epsilon_{m\phi}) + 3H(\epsilon_{\phi}+\epsilon_m + \epsilon_{m\phi} + p_{\phi} + p_m) \simeq 0
\end{align}
which is a continuity equation. The total energy density is seen to be just the sum of the expected matter, scalar and interaction energy density and the pressure likewise.

There is one last, but very handy, relation we can derive in the case where the time-derivatives of the scalar field can be neglected in the Klein-Gordon equation. Starting from $U_{\nabla\phi}$ and using integration by parts we find
\begin{align}\label{ccrel_1}
U_{\nabla\phi} &= -\frac{1}{2}\int d^3r \phi \nabla(f_X\nabla \phi)\nonumber\\
&=-\frac{1}{2}\int d^3r \phi\left(f_{,\overline{\phi}}-f_{,\phi}  + \frac{\beta(\phi)\rho_m}{M_{\rm Pl}}-\frac{\beta(\overline{\phi})\overline{\rho}_m}{M_{\rm Pl}}\right)
\end{align}
Now if $\beta$ is a constant then this equation simplifies to
\begin{align}\label{ccrel}
U_{\nabla\phi} + \frac{1}{2}U_{\log A}&= -\frac{1}{2}\int d^3r \phi\left(f_{,\overline{\phi}}-f_{,\phi}\right)
\end{align}
which can be used separately from the Layzer-Irvine equation as a consistency relation or together with the Layzer-Irvine equation itself to remove e.g. the term $U_{\nabla\phi}$. 

The advantage of using Eq.~(\ref{ccrel_1}) (or Eq.~(\ref{ccrel})) is that it does not depend on time-derivatives and can be used for an arbitrary static configuration. This equation can serve as a novel test of the scalar field solver in an N-body code. The advantage of this test over current static tests is that it allows us to test the code using a realistic density distribution, i.e. one similar to that encountered in numerical simulations. One can also use this relation at each time-step when performing numerical simulations as an accuracy check.

 
\section{Specific models}
In this section we go through specific models and conditions where additional approximations and simplifications can be made. The simplifications we make are those that apply for N-body simulations and are not always applicable in general. We start by checking that the equation we have derived gives predictions that agree with our expectations.

\subsection{Enhanced Gravity}

Lets, as a consistency check, start with the case where we have a fifth-force that has an infinite Compton wavelength and a constant coupling $\beta$. This is achieved by taking $f(X,\phi) = X$ and $A(\phi) = e^{\frac{\beta\phi}{M_{\rm Pl}}}$. This case corresponds to standard gravity, but where Newtons constant $G$ is larger by a factor $1+2\beta^2$. Under the assumption that we can neglect time-derivatives in the Klein-Gordon equation for the scalar field we find
\begin{align}
\log A = \frac{\beta\phi}{M_{\rm Pl}} = 2\beta^2 \Phi_N
\end{align}
giving 
\begin{align}
U_{\log A} = 4\beta^2U_N,~~~~~U_{\nabla\phi} = -2\beta^2U_N
\end{align}
Since $\frac{\beta\phi}{M_{\rm Pl}} = 2\beta^2\Phi_N \ll 1$ we can safely put $A = 1$. This means we can also take $U_A = 0$ and $U_{\dot{\phi}}$ is negligible as this is second order in the time-derivative of the gravitational potential. The term $U_f \equiv 0$ as $g-\overline{g} \equiv 0$ and this also holds if we add a constant potential (a cosmological constant) to the scalar field. This leaves us with the equation
\begin{align}
\frac{\partial}{\partial t}\left(T + U_{\rm tot}\right) + H\left(2T + U_{\rm tot}\right) = 0
\end{align}
where $U_{\rm tot} = U_N(1+2\beta^2)$. This is the correct result as can be seen by making the substitution $G\to G(1+2\beta^2)$ in the original Layzer-Irvine equation Eq.~(\ref{lieq}).

\subsection{Yukawa interaction}\label{yukawa}
The next simplest case is a massive scalar field coupled to matter. This case leads to a total gravitational force between two point masses of the Yukawa type
\begin{align}
\vec{F} = -\frac{GM_1M_2}{r^2}\left(1+2\beta^2(1+mr)e^{-mr}\right)\frac{\vec{r}}{r}
\end{align}
where $2\beta^2$ is the strength and $m^{-1}$ is the range of the matter-scalar interaction.

This scenario is achieved by taking $f(X,\phi) = X - V(\phi)$ where $V(\phi) = \frac{1}{2}m^2\phi^2$ and $A(\phi) = e^{\frac{\beta\phi}{M_{\rm Pl}}}$.

As for the case above we can neglect $U_A $ and $U_{\dot{\phi}}$, but now the term $U_f$ is non-zero
\begin{align}
U_f = \int d^3r \frac{1}{2}m^2\left(\phi^2 - \overline{\phi}^2\right)
\end{align}
and represents the potential energy of the scalar field itself. From Eq.~(\ref{ccrel}) we get the very simple relation
\begin{align}
U_{\nabla\phi} + \frac{1}{2}U_{\log A} + U_f = 0
\end{align}
which gives the Layzer-Irvine equation
\begin{align}\label{lieqyukawa}
\frac{\partial}{\partial t}(T + U_{\rm tot}) + 2H(2T + U_{\rm tot} - 2U_f) = 0
\end{align}
where $U_{\rm tot} = U_N + \frac{1}{2}U_{\log A}$. We can now check that we get the correct value for $U_{\rm tot}$.

If we assume the time-derivatives can be neglected then we can Fourier transform the Klein-Gordon equation with the result
\begin{align}
\mathcal{F}(\phi) = \frac{\beta\mathcal{F}(\delta\rho_m)}{M_{\rm Pl}} \frac{k^2}{k^2+m^2}
\end{align}
Taking the inverse Fourier transform and using the convolution theorem together with $\mathcal{F}^{-1}\left(\frac{4\pi}{m^2+k^2}\right) = \frac{1}{r}e^{-mr}$ we can write down an explicit solution for the scalar field
\begin{align}
\frac{\beta\phi}{M_{\rm Pl}} = -2\beta^2 G \int \frac{\delta\rho(r_1)d^3r_1}{|\vec{r}-\vec{r_1}|}e^{-m|\vec{r}-\vec{r_1}|}
\end{align}
From this it follows that
\begin{align}
U_{\rm tot} &= U_N + \frac{1}{2}U_{\rm log A}\nonumber\\
& = -\frac{G}{2}\int\int\frac{\delta\rho(r_1,t)\delta\rho(r_2,t)d^3r_1d^3r_2}{|\vec{r_1}-\vec{r_2}|}\times\nonumber\\
&~~~~~~~~~~~~~~~ \times (1+2\beta^2 e^{-m |\vec{r_1}-\vec{r_2}|})
\end{align}
which is the correct potential energy for a Yukawa interaction combined with gravity. In the limit $m\to 0$ we recover the case discussed above. Our result Eq.~(\ref{lieqyukawa}) agrees with that of \cite{lieq_4} with the exception of the term $U_f$ which was not taken into account in their phenomenological approach.

\subsection{Non-clustering scalar field}

In theories where the scalar field does not cluster significantly the factor $\frac{\partial \log A}{\partial \log a}$ can be taken to be equal to the background value giving $\delta U_x = \frac{\partial \log \overline{A}}{\partial \log a}U_x$. 

For quintessence models $f=X-V$ and the coupling to matter is zero ($\beta \equiv 0$) giving the same equation as for standard gravity. The modifications from standard gravity are only implicit in the evolution of $H(t)$. This is also expected as the quintessence field only affects the background cosmology.

Coupled quintessence \cite{cq} is a class of models where dark matter and dark energy (given by the scalar field $\phi$) have interactions. General models in this class have a time-varying coupling $\beta(\phi) \simeq \beta(\overline{\phi}) \equiv \beta(a)$. The interaction range in these models, when explaining dark energy, are of the order of the Hubble radius giving $\log A \simeq 2\beta^2(a)\Phi_N$ and the Layzer-Irvine equation simplifies greatly to
\begin{align}\label{lieq_cc}
&\frac{\partial}{\partial t}\left(T + U_{\rm tot}\right) + H\left(2T + U_{\rm tot}\right) + \frac{\beta(a)}{M_{\rm Pl}}\dot{\overline{\phi}}(T-2U_{\rm tot}) = 0
\end{align}
where $U_{\rm tot} = (1+2\beta^2(a)) U_N$ is the total potential energy. This equation agrees\footnote{In the notation of \cite{lieq_wang} we have $\zeta_1 = \frac{1}{3}\frac{\beta(a)}{M_{\rm Pl}}\frac{d\phi(a)}{d\log a}$ and $\zeta_2=0$ for the model considered here. Inserting this in their Eq.~(17) gives our Eq.~(\ref{lieq_cc}). Likewise, by comparing our notation with that of \cite{lieq_new} we find $\overline{\zeta} = \frac{\beta(a)}{M_{\rm Pl}}\frac{d\phi(a)}{d\log a}$ which in their Eq.~(5) gives our Eq.~(\ref{lieq_cc}.} with the result found in \cite{lieq_new} and \cite{lieq_wang}.
 
 \subsection{Chameleon-like theories}

Chameleon-like modified gravity theories refers to models given by the action Eq.~(\ref{action}) with $f=X-V(\phi)$ where the effective potential $V_{\rm eff} \equiv V(\phi) + A(\phi)\rho_m$ has a minimum $\phi_{\rm min}(\rho_m)$ and where the mass $m^2(\phi) = V_{\rm eff,\phi\phi}$ at this minimum is an increasing function of $\rho_m$. Examples of such model are the $f(R)$/chameleon \cite{chameleon}, symmetron \cite{symmetron} and environmental dependent dilaton \cite{env_dilaton}. In these models local gravity constraints forces $\frac{\partial \log A}{\partial \log a} \ll 1$ \cite{unified_winther} and all the terms $\delta U_x$ can be neglected. This also generally implies that $|\dot{\phi}| \ll |\vec{\nabla}\phi|$ implying $U_{\dot{\phi}} \ll U_{\nabla\phi}$, an approximation often refereed to as the quasi-static approximation \cite{symmetron_winther} and is the reason why N-body simulation of these theories can neglect the time-derivatives in the Klein-Gordon equation\footnote{Recently, a new code came out where the full Klein-Gordon equation is solved for the first time in an N-body code \cite{mota_claudio}.}. This leaves us with the simplified equation
\begin{align}\label{fofr_eq}
&\frac{\partial}{\partial t}\left(T + U_N + U_{\log A} + U_{\nabla\phi} + U_f + U_A\right) \nonumber\\
&+ H\left(2T + U_N - U_{\nabla\phi}-3U_f\right) = 0
\end{align}

\section{Implementation in N-body codes}\label{sec4}
In this section we discuss how to numerically implement the modified Layzer-Irvine equation in an N-body code and how we can monitor the level of which it is satisfied.

For standard gravity the kinetic energy of the dark matter particles is given by
\begin{align}
T =\int d^3r \frac{1}{2}v^2\rho_m = \sum_{i=1}^{N_{\rm part}} \frac{1}{2}m_i v_i^2
\end{align}
where $m_i$ is the mass of each N-body particle with $m_i = \frac{\rho_{m0}B_0^3}{N_{\rm part}}$ when all particles have the same mass. $B_0$ denotes the boxsize at $a=1$ and $N_{\rm part}$ the number of particles in the simulation.

Using the Poisson equation and integration by parts, the gravitational potential energy can be written
\begin{align}
U_N &=\int d^3r \frac{1}{2} \Phi_N \delta\rho_m = \frac{1}{4\pi G}\int d^3r \frac{1}{2} \Phi_N \nabla_r^2\Phi_N\\
&=-\frac{1}{8\pi G}\int d^3r (\vec{\nabla}_r\Phi_N)^2
\end{align}
In an N-body code we can approximate this potential (here for a grid based code) by
\begin{align}
U_N \simeq -\frac{1}{8\pi G}\sum_{i=0}^{N_{\rm cell}}dr_{\rm cell~i}^3 (\vec{F}_N)_i^2
\end{align}
where the sum is over all the cells of the grid structure, $(\vec{F}_N)_i = -(\vec{\nabla}_r\Phi_N)_i$ is the force field and $dr_{\rm cell ~ i}^3$ is the volume of grid-cell $i$. Note that the gradient and the volume element is in terms of the physical variable: $\nabla_r = \frac{1}{a}\nabla_x$ and $d^3r =a^3 dx^3$.


In modified gravity, the kinetic energy is modified compared to standard gravity as the mass of the particles are now $\phi$ dependent
\begin{align}
T = \sum_{i=1}^{N_{\rm part}} \frac{1}{2}m_i(\phi) v_i^2
\end{align}
where $m_i(\phi) = A(\phi) m_i$ with $\sum_i m_i = \rho_{m0} B_0^3$. As for standard gravity we have
\begin{align}
U_N \simeq -\frac{1}{8\pi G}\sum_{i=0}^{N_{\rm cell}}dx_{\rm cell~i}^3 (\vec{F}_N)_i^2
\end{align}
The fifth-force potential can be rewritten using the Poisson equation and integration by parts to give
\begin{align}
U_{\log A} &= -\frac{1}{8\pi G}\int d^3r~2(\vec{\nabla}_r\Phi_N) \cdot (\vec{\nabla}_r\log A)
\end{align}
which can be evaluated as
\begin{align}
U_{\log A} \simeq -\frac{1}{8\pi G}\sum_{i=0}^{N_{\rm cell}}dx_{\rm cell~i}^3 2(\vec{F}_N)_i\cdot(\vec{F}_\phi)_i
\end{align}
where $(\vec{F}_N)_i = -\left(\frac{\beta(\phi)}{M_{\rm Pl}}\vec{\nabla}_r\phi\right)_i$ is the fifth-force in grid cell $i$. The other potentials are trivial to calculate, for example
\begin{align}
U_A &\simeq \sum_{i=0}^{N_{\rm cell}}dx_{\rm cell~i}^3 (A(\phi_i) - A(\overline{\phi}))
\end{align}
There is also a further simplification for theories with constant coupling $\beta$ (i.e. $\log A \equiv \frac{\beta\phi}{M_{\rm Pl}}$) where we can write the term $U_{\nabla\phi}$ as 
\begin{align}
U_{\nabla\phi} \simeq +\frac{(2\beta^2)^{-1}}{8\pi G}\sum_{i=0}^{N_{\rm cell}}dx_{\rm cell~i}^3 (\vec{F}_\phi)_i^2
\end{align}

When implementing the Layzer-Irvine equation in an N-body code it is convenient to work with the normalized potentials
\begin{align}
E_i \equiv \frac{a^2 U_i}{(H_0B_0)^2\rho_{m0}B_0^3}
\end{align}
In this form the potentials are dimensionless and also the kinetic friction term $2HT$ is removed from the equation. This is also the definition used in {\tt{RAMSES}} \cite{ramses}, for which the N-body code {\tt ISIS} \cite{isis} we have used to implement these equations, is based on.

To define the deviation from the modified Layzer-Irvine equation we first start by writing it as
\begin{align}
\sum_i \left(\alpha_i\frac{\partial}{\partial t}  + \gamma_i H\right)E_i = 0
\end{align}
where $\alpha_i$ and $\gamma_i$ are constants or functions of the background cosmology only. In order to evaluate this equation numerically, it is more convenient to rephrase it as the integral equation
\begin{align}\label{leqn}
\sum_i \alpha_i\left(E_i(a_j) - E_i(a_0) \right) + \int_{a_0}^{a_j}\sum_i (\gamma_i E_i)\frac{da}{a} = 0
\end{align}
We denote the left hand side of the equation above as $\sigma_j$. To have something to compare $\sigma_j$ against we define
\begin{align}
\Sigma_j \equiv \sum_i |\alpha_i|\left(|E_i(a_j)| - |E_i(a_0)| \right) + |\int_{a_0}^{a_j}\sum_i (\gamma_i E_i)\frac{da}{a}|
\end{align}
We can now define the error, or deviation, from the Layzer-Irvine equation at time-step $j$ by
\begin{align}
\epsilon(a_j) \equiv \frac{\sigma_j}{\Sigma_j}
\end{align}
The function $\epsilon(a)$ will be referred to as the Layzer-Irvine constant.

It only remains to define how we calculate the integral in Eq.~(\ref{leqn}). In an N-body code we only have the potentials $E_i(a_j)$ at each discrete time-step $j$ and must therefore use some approximation for the integral. We start by writing the integral in Eq.~(\ref{leqn}) as
\begin{align}
I_j = \int_{a_0}^{a_j}\sum_i (\gamma_i E_i)\frac{da}{a} = \sum_{k=1}^j \int_{a_{k-1}}^{a_k}\sum_i (\gamma_i E_i)\frac{da}{a}
\end{align}
so that $I_j = I_{j-1} + \delta I_j$ where 
\begin{align}
\delta I_j = \int_{a_{j-1}}^{a_j}\sum_i (\gamma_i E_i)\frac{da}{a}
\end{align}
This integral is approximated by the mean value of the discrete integrand and an exact integration of $\int da/a$ giving
\begin{align}
\delta I_j \simeq \frac{\left[\sum_i (\gamma_i E_i)\right]_{a=a_{j-1}}+\left[\sum_i (\gamma_i E_i)\right]_{a=a_{j}}}{2}\log\left(\frac{a_j}{a_{j-1}}\right)
\end{align}

\section{Tests on N-body Simulations}\label{sec5}
We have run N-body simulations of modified gravity models to see whether the Layzer-Irvine equation developed here is satisfied and also to see what level of violation we would get if a mistake is made in the numerical implementation. For all the modified gravity models we present tests of here we have beforehand tested the code against static configurations where known analytic solutions exist and found a good agreement. We will therefore assume that the implementation of the (static) Klein-Gordon equation is correct and the tests we perform will tell us if the code is able to accurately solve for the time-integration of these models.

The N-body simulations performed in this paper is done with the {\tt ISIS} code \cite{isis} which is based on the public available code {\tt RAMSES} \cite{ramses}.


\subsection{Enhanced gravity and the Yukawa interaction}

We have implemented the Yukawa interaction model described in Sec.~(\ref{yukawa}) in the N-body code ISIS \cite{isis}. We ran simulations in a box of size $B_0= 200~{\rm Mpc}/h$ with $N=128^3$ particles and a standard WMAP7 cosmology starting from $z=20$. The model parameters used in this test are $m^{-1} =\left\{1, 5,\infty\right\}~{\rm Mpc}/h$ together with $2\beta^2 = \left\{0.01,0.1,0.5\right\}$. The $m^{-1} = \infty$ run is equivalent to standard gravity with an enhanced gravitational constant $G\to G(1+2\beta^2)$ and serves as a benchmark for the modified gravity models we will look at below.

\begin{figure}
\includegraphics[width=\columnwidth]{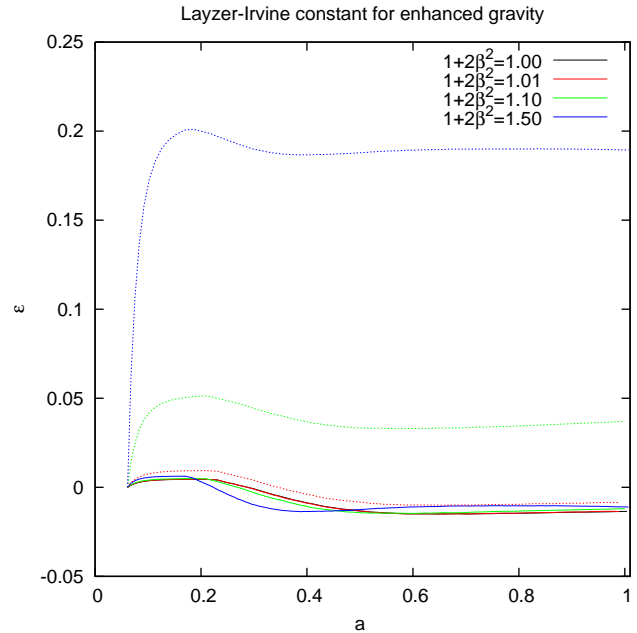}
\caption{The Layzer-Irvine constant as function of scale factor for the enhanced gravity model (solid lines) $G_{\rm eff} = G(1+2\beta^2)$. The dotted lines show the corresponding Layzer-Irvine constant calculated using the pure GR equation Eq.~(\ref{lieq}), i.e. when not taking the potential energies of the scalar field ($U_{\nabla\phi}$ and $U_{\log A}$) into account.}
\label{fig1}
\end{figure}

In Fig.~(\ref{fig1}) we show the Layzer-Irvine constant $\epsilon$ for the enhanced gravity model ($m^{-1}=\infty$) with $1+2\beta^2 = 1.5$, $1+2\beta^2 = 1.1$, $1+2\beta^2 = 1.01$ and standard gravity $\beta=0$. All the simulations use the same initial conditions and the same background cosmology. We find that $\epsilon \lesssim 0.01$ during the whole evolution for all runs which is also what we get for the standard gravity simulation. This test tells us that even when gravity is enhanced the code is still able to accurately solve the N-body equations.

The dotted line in Fig.~(\ref{fig1}) shows the Layzer-Irvine constant calculated using the Layzer-Irvine equation for standard gravity Eq.~(\ref{lieq}). This result is equivalent to what we would get if we made a mistake in the numerical implementation consisting of taking the pre-factor in the geodesic equation to be a factor $1+2\beta^2$ larger than the correct value. The huge deviation we see, even for $1+2\beta^2=1.1$, demonstrates the usefulness of the Layzer-Irvine equation: a small mistake in the numerical implementation of the geodesic equation will show up as a clear violation in the Layzer-Irvine constant.

\begin{figure*}
\includegraphics[width=\columnwidth]{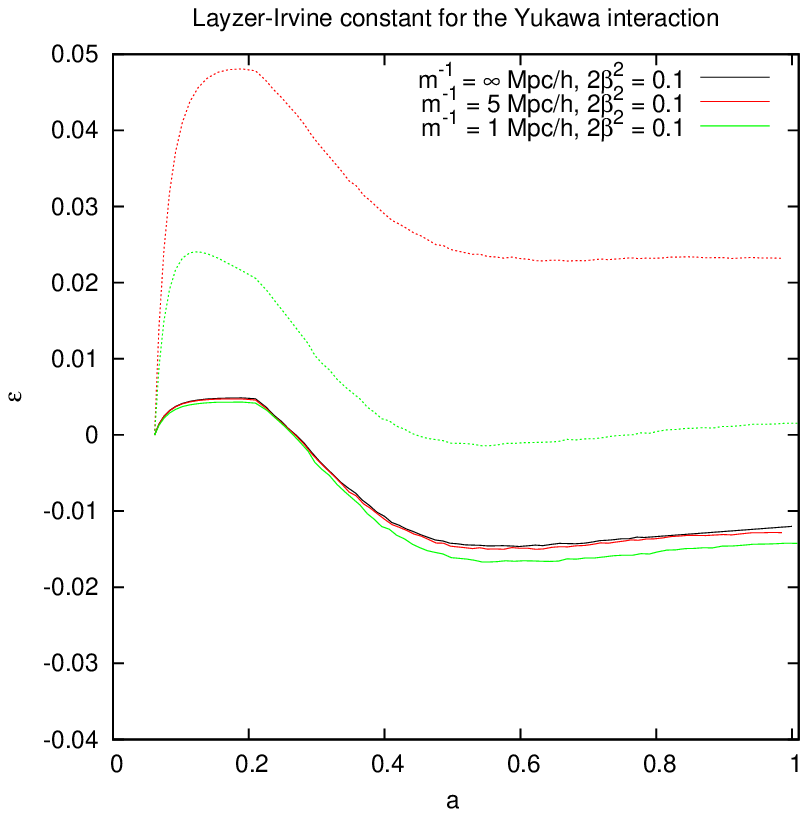}
\includegraphics[width=\columnwidth]{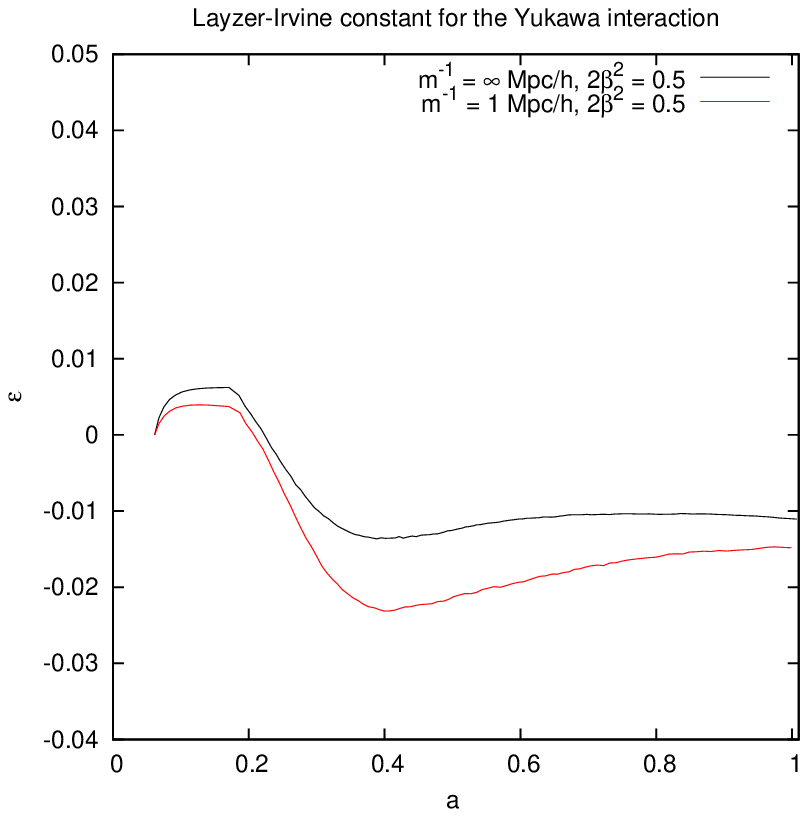}
\caption{The Layzer-Irvine constant as function of scale factor for Yukawa interaction model (solid lines) with coupling strength $2\beta^2 = 0.1$ (left) and $2\beta^2 = 0.5$ (right). The dotted lines show the corresponding Layzer-Irvine constant calculated using the pure GR equation Eq.~(\ref{lieq}).}
\label{fig2}
\end{figure*}

In Fig.~(\ref{fig2}) we show the Layzer-Irvine constant for the Yukawa model with $2\beta^2 = 0.1$ and $m^{-1}=\{ 1,5,\infty\}$ Mpc/h together with an enhanced gravity simulation with the same strength. The Layzer-Irvine constant is just as well satisfied for the Yukawa simulations as for the pure gravity simulation.

\begin{figure}
\includegraphics[width=\columnwidth]{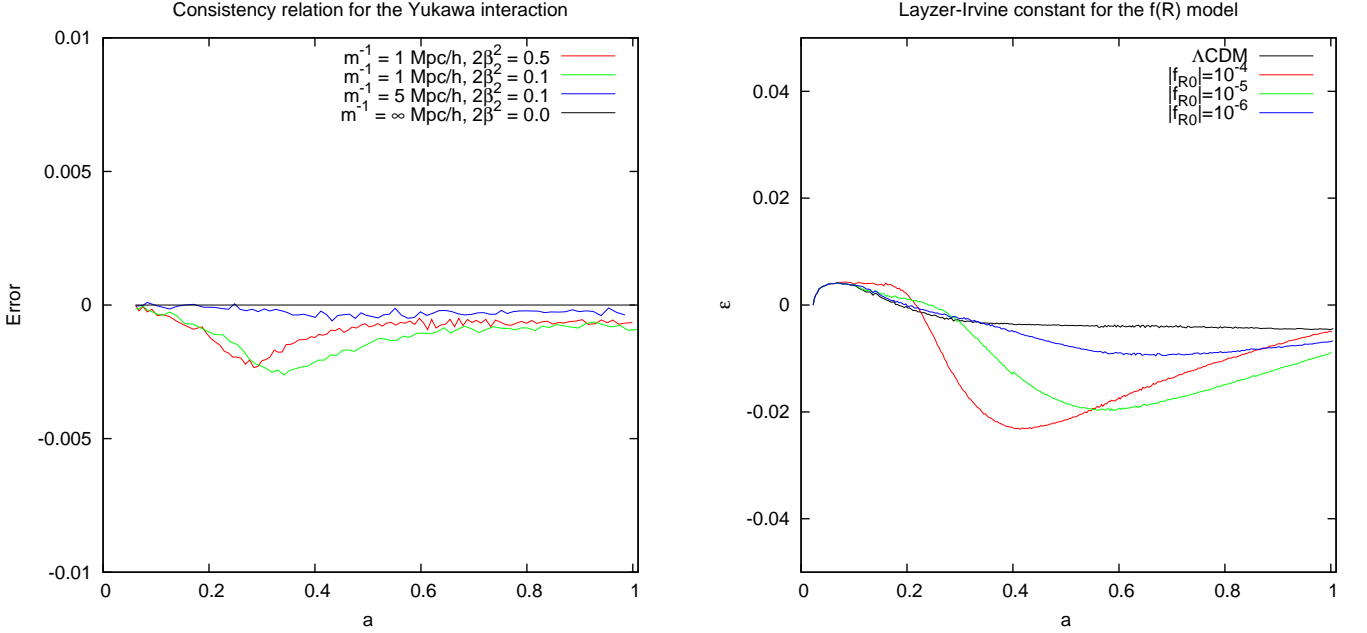}
\caption{Test of the relation $U_{\nabla\phi} + \frac{1}{2}U_{\log A} + U_f \equiv 0$ for the Yukawa interaction model. The error is defined as $(U_{\nabla\phi} + \frac{1}{2}U_{\log A} + U_f ) / (|U_{\nabla\phi}| + \frac{1}{2}|U_{\log A}| + |U_f|)$.}
\label{fig3}
\end{figure}

For the Yukawa interaction we also test the relation Eq.~(\ref{ccrel}). This relation does not involve time evolution so the results in one time-step is independent of the previous time-steps and this allow us to use it to test the code for a realistic\footnote{With realistic we mean a density distribution similar to what we encounter when performing numerical simulations.} static configuration where no analytical solutions can be found. The results are shown in Fig.~(\ref{fig3}). The deviation from this relation (measured against the sum of the absolute values of the three terms) for the most extreme model are found to be less than $0.2\%$ during the whole evolution.

In all cases we see that the Layzer-Irvine constant for the Yukawa interaction is small and the deviation we find is roughly the same as for the enhanced gravity simulation with the same $\beta$. 

We note that the (small) violation of the Layzer-Irvine equation is closely related to the creation of new refinements in the code. The relative fraction of new refinements being created in the simulations peaks during the period $0.2\lesssim a\lesssim 0.5$ which agrees with the time when we see the largest deviation. This happens because when new refinement are created we automatically increase the accuracy in the calculation of the potentials while leaving the kinetic energy (which comes from the particles) untouched. We also note that the evolution of the Layzer-Irvine constant for any model, standard gravity included, depends sensitively on the refinement criterion, the number of particles and the time-stepping criterion used in the simulation. A complete study of all these effects are beyond the scope of this paper.

\begin{figure}
\includegraphics[width=\columnwidth]{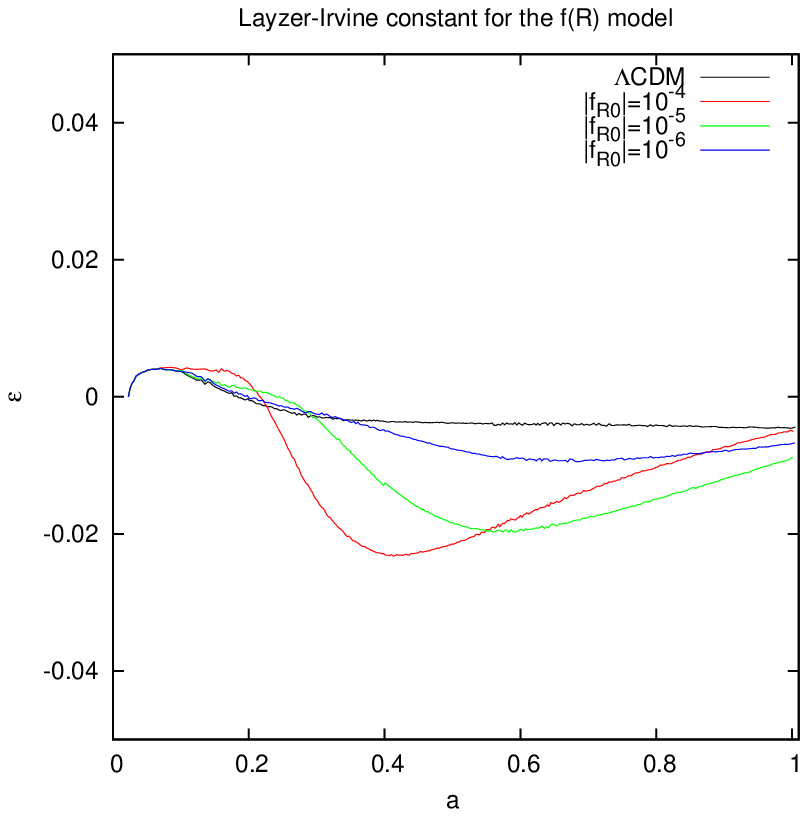}
\caption{The Layzer-Irvine constant as function of scale factor for the $f(R)$ simulations in \cite{isis}.}
\label{fig4}
\end{figure}

\subsection{$f(R)$ gravity}

$f(R)$-gravity can be written as a scalar tensor theory where $A(\phi) = e^{\frac{\beta\phi}{M_{\rm Pl}}}$ with $\beta = 1/\sqrt{6} \approx 0.408$ and for some model specific potential $V(\phi)$ \cite{brax_fofr}.

The particular Hu-Sawicky $f(R)$ model \cite{fofr_hu} has been implemented in {\rm ISIS}. The implementation has been properly tested against analytical (static) configurations and against results from the literature. The code was found to work accurately.

For the simulations performed in \cite{isis} we have calculated the Layzer-Irvine constant using Eq.~(\ref{fofr_eq}) which is consistent with the approximations used in the simulation. These simulations all have $N=512^3$ particles in a box of size $B_0 = 256~{\rm Mpc}/h$ using a standard WMAP7 cosmology. See \cite{isis} for more details.

In Fig.~(\ref{fig4}) we show the Layzer-Irvine constant for the three simulations with the model parameter $|f_{R0}| = \{10^{-4},10^{-5},10^{-6}\}$ compared to a $\Lambda CDM$ simulation using the same initial conditions. For a more complete description of the Hu-Sawicky model see for example \cite{fofr_hu,fofr_baojiu}. 

We find that the Layzer-Irvine constant has a maximum deviation of $\sim 2\%$ which is comparable with the evolution of the Yukawa interaction with $\beta=0.5$ presented above. 

\section{Conclusions}\label{sec6}
We have derived the Layzer-Irvine equation, describing quasi-Newtonian energy conservation for a collisionless fluid in an expanding background, for a large class of scalar-tensor modified gravity theories. The equation derived have been tested in N-body simulations of modified gravity theories.

Monitoring the Layzer-Irvine equation is one of the few tests that directly probes the time-evolution of a simulation.

We demonstrated that a mistake made in the implementation of a modified gravity theory, consisting of a wrong pre-factor in the geodesic equation off by no more than a few percent from the correct one, will lead to a huge violation of the Layzer-Irvine equation. Such a mistake will also give effects on the matter power-spectrum, but these can be degenerate with cosmic variance.

As a test, the Layzer-Irvine equation can be used in several different ways. When implementing new models in an N-body code one often make several approximations to simplify the equations of motion. One way to apply it is to take the actual equation we put into the code, derive the corresponding Layzer-Irvine equation and run the simulation. The results from this equation will tell us how good the code solves the equations we actually try to solve, i.e. how good is the accuracy and the methods used. Secondly, we can take the full Layzer-Irvine equation and test it. The results from this equation can tell us something about how good the approximations we have used are. Lastly, we have shown how the relation Eq.~(\ref{ccrel_1}) can be used as a new static test which can be applied to any density distribution where no analytic or semi-analytic solution of the Klein-Gordon can be found.

There are scalar-tensor theories that are not covered by our analysis, like for example the Galileon, however the same methods we used here can easily be applied to any scalar field theory of interest.

\section*{Acknowledgements}
The author is supported by the Research Council of Norway FRINAT grant 197251/V30. I would like to thank David Mota for many useful discussions about this topic.

\end{document}